\documentclass[prb,twocolumn,amsmath,amssymb,amsfonts]{revtex4}%
\usepackage{graphicx}
\usepackage{dcolumn}
\usepackage{bm}
\usepackage{amsmath}
\usepackage{amsfonts}
\usepackage{amssymb}%
\begin{document}
\title{Keldysh study of point-contact tunneling between superconductors}
\author{C.~J.~Bolech}
\author{T.~Giamarchi}
\affiliation{Universit\'{e} de Gen\`{e}ve, DPMC, 24 Quai Ernest Ansermet, CH-1211
Gen\`{e}ve 4, Switzerland}
\date{June 6$^{\text{th}}$, 2004}

\begin{abstract}
We revisit the problem of point-contact tunnel junctions involving
one-dimensional superconductors and present a simple scheme for computing the
full current-voltage characteristics within the framework of the
non-equilibrium Keldysh Green function formalism. We address the effects of
different pairing symmetries combined with magnetic fields and finite
temperatures at arbitrary bias voltages. We discuss extensively the importance
of these results for present-day experiments. In particular, we propose ways
of measuring the effects found when the two sides of the junction have
dissimilar superconducting gaps and when the symmetry of the superconducting
states is not the one of spin-singlet pairing. This last point is of relevance
for the study of the superconducting state of certain organic materials like
the Bechgaard salts and, to some extent, for ruthenium compounds.

\end{abstract}
\maketitle

\section{Introduction\label{sec:intro}}

The theory of superconductivity by Bardeen, Cooper and Schrieffer (BCS) is one
of the most important achievements of condensed matter theory. Some of the
most striking consequences of this theory concern the tunneling to and from a
superconductor. Indeed, the history of tunneling experiments and applications
is strongly linked to that of superconductivity. Not only some of the most
crucial experimental verifications of the BCS theory came from tunneling
experiments,\cite{wolf1989} but also some of the most important practical
applications of superconductivity involve Josephson tunneling junctions. To
describe the manifold of experimental and practical situations, two limiting
cases are usually considered: planar interfaces and point contacts. As of
latter, point-contact tunneling \textit{per se} acquired renewed relevance
with the development of scanning tunneling microscopy
(STM),\cite{quate1986,binnig1987,binnig1999} that is today at the forefront of
the experimental techniques used to study unconventional superconductors. For
STM the tip can be modeled using some idealized geometry. For example the
cases of spherical,\cite{tersoff1983,tersoff1985}
conical\cite{suderow2000,suderow2002} and pyramidal\cite{cuevas1998} tip
geometries were considered in the literature (the last two were used to model
very close STM contacts). The point-contact approximation is therefore the
simplest one for the kind of tunneling processes that take place on STM
experiments. Other ways to realize point contacts include the use of break
junctions and pressed crossed wires.

The simplest theoretical models used to interpret experiments involving
superconducting tunneling are typically based on a simple scattering picture
and go generally under the name of \textit{semiconducting band models}%
.\cite{nicol1960,klapwijk1982,blonder1982,octavio1983} A more systematic
approach is that based on the tunneling
Hamiltonian.\cite{cohen1962,wilkins1969,cuevas1996} A large series of recent
experiments\cite{scheer1997,scheer1998,ludoph2000,scheer2001,rubio2003,hafner2004}
on atomic-size contacts showed impressive agreement with the theory;
achieving, some of them, detailed microscopic description of the contacts. The
current transport in these systems can be described as taking place through a
small number of independent \textit{conduction channels}, each of them well
described by a point-contact model. In some experiments even the observation
of single-channel transport was possible.

Tunneling can thus be used as a very efficient probe of the properties of the
leads. In particular one can expect to use it to determine the symmetry of the
superconducting order parameter in the leads. However, the previous
theoretical analysis of point-contact tunneling, although efficient in simple
cases, are too cumbersome to be easily generalized to more complex cases such
as unconventional order parameters at finite temperatures and finite magnetic
fields. Simplified semiclassical methods
exist,\cite{millis1988,cuevas2001,kopu2004} but they suffer from their own
limitations, for instance when one is dealing with anisotropic
superconductors.\cite{kopnin2001} Thus a general and simple microscopic theory
of point-contact tunneling was clearly lacking, and is necessary in order to
take into account some of the complications of unusual superconductivity.
Providing such a theory is the purpose of this work. We use a Keldysh
formalism, to be able to compute the full current-voltage characteristics and
gain access to the effects of external magnetic fields, potential scattering
barriers and finite temperatures on the transport properties of different
junctions at arbitrary finite voltages. Contrarily to previous implementations
of this technique, using the solution of difference
equations,\cite{cuevas1996} we here obtain and diagonalize the full tunneling
action for the point contact tunneling junctions involving normal-metal and
superconductor leads. This allows one to easily incorporate complications such
as triplet pairing in the leads, finite temperature and finite magnetic field.
We explore in particular the physical properties of tunneling systems with
leads with triplet paring parameters.

Indeed, although the possibility of having triplet pairing was
investigated\cite{anderson1961,balian1963} soon after the BCS theory, and such
an unconventional scenario was found about the same time in the \textit{p}%
-wave spin-triplet superfluid state of $^{3}$\textrm{He}%
,\cite{osheroff1972a,osheroff1972b,anderson1975} the quest to identify a
\textit{p}-wave charged superfluid proved much more challenging. A class of
candidates for triplet pairing, though the evidence is as yet not completely
conclusive, are the organic superconductors\cite{chemicalreview,ishiguro2002}
and the ruthenates.\cite{rice1995,baskaran1996} There are also proposals of
spin-triplet pairing phases for some heavy fermion superconductors like
\textrm{UPt}$_{3}$, but the issue remains more open in those
cases.\cite{joynt2002} The organic compounds are the most interesting for us
due to the quasi-one-dimensional nature of their normal phases, and also
because there is currently considerable debate on the symmetry of the
superconducting phase.\cite{ISCOM2003,oh2004,joo2004} For the ruthenates, on
the other hand, the triplet pairing seems already backed up by a considerable
amount of experimental evidence.\cite{rice2001,mackenzie2003} To sort out this
question of the symmetry of the order parameter, tunneling can thus be an
invaluable tool. Recently, STM tunneling experiments were used to study the
symmetry of the superconducting phase of \textrm{Sr}$_{2}$\textrm{RuO}$_{4}$
and other compounds. No such attempts were as yet made in the case of the
quasi-one-dimensional organic salts, but efforts in this direction are on
their way. Also recently, preliminary experiments involving junctions between
two (different) Bechgaard salts were performed, and they showed a number of
puzzling features including a zero-bias conductance peak (\textquotedblleft
anomaly\textquotedblright) and zero excess current.\cite{ha2003} In that
context, a precise theory of the particularities of point-contact tunneling
involving spin-triplet superconductors, done in the microscopic framework of
the tunneling Hamiltonian models, was called for. Given the nature of these
systems, it is difficult to perform phase sensitive experiments such as the
ones that were, for the cuprates, smoking guns to fix the symmetry of the
order parameter. We show, however, that the tunneling spectrum has
characteristic features, such as the magnetic field dependence, that can be
used to unambiguously determine the order parameter symmetry in these systems.
A short account of part of the results of this paper was published
previously.\cite{bolech2004}

The rest of the paper is organized as follows. In
Sec.~\ref{sec:model} we present the model that we use to describe
the point-contact junction geometry between either normal or
superconducting leads. In Sec.~\ref{sec:lap} we work out a way of
finding the tunneling characteristics using a non-equilibrium
(Keldysh) formalism. This allows us to obtain the current-voltage
characteristics for arbitrary voltage, temperature or magnetic
field, for junctions with either normal, or singlet or triplet
superconducting leads. All technical details have been confined to
these two sections, while the two remaining sections deal with the
physical consequences of our findings. Readers only interested in
those can thus safely jump to Sec.~\ref{sec:tunchar}, where the
physics of such junctions is discussed in detail. Those results
are put in context within different experimental possibilities in
Sec.~\ref{sec:experiments}. In particular we discuss there the
possibility of using tunneling experiments to probe the nature of
the superconducting pairing in the organic superconductors. In
Sec.~\ref{sec:summary} we close the paper with a general
discussion of the implications of our results.

\section{Model of the Point-contact Junction\label{sec:model}}

Using a non-equilibrium Keldysh formalism we calculate the full
current-voltage characteristics of different types of tunnel junctions where
each side of the junction can be either a normal metal (N), a singlet (S) or a
triplet (T) superconductor. We start from a tunneling Hamiltonian
formulation,
\begin{align}
H &  =H_{1}+H_{2}+H_{\mathrm{tun}}\\
H_{\mathrm{tun}} &  =\sum_{\ell,\ell^{\prime},\sigma}t_{\ell\ell^{\prime}%
}~\psi_{\ell\sigma}^{\dagger}\left(  0\right)  \psi_{\ell^{\prime}\sigma}^{%
%TCIMACRO{\TeXButton{t}{\phantom{\dagger}}}%
%BeginExpansion
\phantom{\dagger}%
%EndExpansion
}\left(  0\right)
\end{align}
The first two terms describe the two leads of the junction (superconducting or
otherwise) and the third one models the tunneling processes in which an
electron with spin $\sigma$ hops from lead $\ell^{\prime}$ into lead $\ell$.
The tunneling matrix is
\begin{equation}
t_{\ell\ell^{\prime}}=%
\begin{pmatrix}
V_{1} & t^{\ast}\\
t & V_{2}%
\end{pmatrix}
\end{equation}
The diagonal terms, $V_{n}$, are local contact potential terms included for
the sake of generality\cite{affleck2000} and the off-diagonal ones are the
tunneling matrix elements taken to be constant consistently with the
assumption of a point contact. Since the number of particles in each lead is a
conserved quantity in the absence of tunneling, we can define the current as
proportional to the rate of change in the relative particle number and
write\cite{cohen1962}%
\begin{equation}
I=\frac{e}{2}\left\langle \partial_{t}\left(  N_{2}-N_{1}\right)
\right\rangle =\frac{e}{2i}\left\langle \left[  H_{\mathrm{tun}},N_{1}%
-N_{2}\right]  \right\rangle ~\text{.}\label{eq:currentdef}%
\end{equation}
Notice that the diagonal part of the tunneling matrix conserves particle
numbers and will not contribute to the current.

To model the superconducting leads in calculations intended to capture the
main features of point-contact transport on conventional superconductors, very
simple models suffice to achieve even quantitative agreement with the
experiment. Contrary to the case in some planar junction experiments,
dimensionality plays little or no role in the tunneling. Therefore one can use
one-dimensional leads to carry out all the standard calculations. The
situation becomes more complex in the case of unconventional superconductors,
mainly because the anisotropic nature of the pair wave-function has to be
taken into account when modeling the leads. The most conspicuous case is that
of the cuprate compounds, for which the putative \textit{d}-wave paring cannot
be modeled within a single-band one-dimensional lead. On the other hand, the
organic superconductors that we are interested in are supposed to have
\textit{p}-wave symmetry. Since both \textit{s}-wave and \textit{p}-wave
symmetries can be modeled in single-band one-dimensional chains, we can
conveniently set up a formalism that encompasses the two cases, as well as the
normal state. In the following, we will consider a one-dimensional band with
two Fermi points and expand the fermion fields around them in the conventional
way,\cite{giamarchi2004}
\begin{equation}
\psi_{\sigma}^{%
%TCIMACRO{\TeXButton{t}{\phantom{\dagger}}}%
%BeginExpansion
\phantom{\dagger}%
%EndExpansion
}\left(  x\right)  \approx e^{-ik_{F}x}\psi_{L\sigma}^{%
%TCIMACRO{\TeXButton{t}{\phantom{\dagger}}}%
%BeginExpansion
\phantom{\dagger}%
%EndExpansion
}\left(  x\right)  +e^{ik_{F}x}\psi_{R\sigma}^{%
%TCIMACRO{\TeXButton{t}{\phantom{\dagger}}}%
%BeginExpansion
\phantom{\dagger}%
%EndExpansion
}\left(  x\right)
\end{equation}
thus defining left and right moving fields (lead indexes were omitted here).
Using these fields and in the spirit of the BCS theory, we introduce the
following four gap functions:
\begin{equation}
\Delta_{a}\left(  x\right)  =\lambda_{a}~\left\langle \alpha~\psi
_{L\bar{\alpha}}^{%
%TCIMACRO{\TeXButton{t}{\phantom{\dagger}}}%
%BeginExpansion
\phantom{\dagger}%
%EndExpansion
}\left(  x\right)  ~\sigma_{\alpha\beta}^{a}~\psi_{R\beta}^{%
%TCIMACRO{\TeXButton{t}{\phantom{\dagger}}}%
%BeginExpansion
\phantom{\dagger}%
%EndExpansion
}\left(  x\right)  \right\rangle
\end{equation}
where Greek indexes are summed over, $a=0,\ldots,3$ and $\sigma_{\alpha\beta
}^{0}$ is the identity matrix while the other three are the usual Pauli
matrices. We use the notation $\bar{\alpha}=-\alpha$ with $\alpha\in\left(
\downarrow,\uparrow\right)  \equiv\left(  -1,+1\right)  $.\footnote{Notice
that the object $\alpha\psi_{L\bar{\alpha}}$ transforms as the complex
conjugate representation of the fundamental representation of \textrm{SU(2)}.
In other words, it has the same transformation properties as $\psi_{L\alpha
}^{\dagger}$.} The constants $\lambda_{a}$ would depend on the details of the
microscopic pairing mechanism about which we make no
assumptions.\cite{ISCOM2003} With this definition $\Delta_{0}\left(  x\right)
$ is the spin-singlet order parameter, as in conventional superconductors, and
the other three functions form a vector of spin-triplet order
parameters,\cite{mackenzie2003} $\vec{\Delta}\left(  x\right)  =\Delta\left(
x\right)  \hat{d}\left(  x\right)  $. We use the approximation of dropping the
spatial dependence in the order parameter and, directly in Fourier space, we
write the Hamiltonian for any of the two leads as
\[
K=\xi_{ck\sigma}\psi_{ck\sigma}^{\dagger}\psi_{ck\sigma}^{%
%TCIMACRO{\TeXButton{t}{\phantom{\dagger}}}%
%BeginExpansion
\phantom{\dagger}%
%EndExpansion
}-\left\{  \Delta_{a}\left[  \psi_{Rk\beta}^{\dagger}~\sigma_{\beta\alpha}%
^{a}~\alpha~\psi_{L\bar{k}\bar{\alpha}}^{\dagger}\right]  +\mathrm{h.c.}%
\right\}
\]
where $K=H-\mu N$ with $\mu$ the chemical potential of that lead. All the
indexes are summed over, in particular $c\in\left(  L,R\right)  \equiv\left(
-1,+1\right)  $ sums over the two possible chiralities and $\xi_{ck\sigma
}=cv_{\mathrm{F}}k-\mu-\sigma h$ are the corresponding linear dispersions,
shifted by the inclusion of chemical potential and magnetic field along the
$\hat{z}$-axis (for convenience we will take $v_{\mathrm{F}}=1$). This is the
natural extension to the triplet case of the usual pairing-approximation
Hamiltonian found in BCS theory, we remark that the fact it does not conserve
particle number is an artifact of the anomalous mean field approximation
behind its derivation and has no bearing in the operator definition of the current.

\section{Local Action Approach\label{sec:lap}}

Within the extended-BCS framework, the Hamiltonian remains a quadratic form
including `anomalous' terms. To be able to write it down as a canonical
quadratic form we introduce the following spinor notation:
\begin{equation}
\Psi_{kn\sigma}^{%
%TCIMACRO{\TeXButton{t}{\phantom{\dagger}}}%
%BeginExpansion
\phantom{\dagger}%
%EndExpansion
}\left(  \varpi\right)  =%
\begin{pmatrix}%
%TCIMACRO{\TeXButton{s}{\phantom{\sigma}}}%
%BeginExpansion
\phantom{\sigma}%
%EndExpansion
~\psi_{Rk\sigma}^{%
%TCIMACRO{\TeXButton{t}{\phantom{\dagger}}}%
%BeginExpansion
\phantom{\dagger}%
%EndExpansion
}\left(  \varpi\right)  \\
\sigma~\psi_{L\bar{k}\bar{\sigma}}^{\dagger}\left(  \bar{\varpi}\right)
\end{pmatrix}
\equiv%
\begin{pmatrix}%
%TCIMACRO{\TeXButton{-}{\phantom{-}}}%
%BeginExpansion
\phantom{-}%
%EndExpansion
\psi_{Rk\uparrow}^{%
%TCIMACRO{\TeXButton{t}{\phantom{\dagger}}}%
%BeginExpansion
\phantom{\dagger}%
%EndExpansion
}\left(  \varpi\right)  \\%
%TCIMACRO{\TeXButton{-}{\phantom{-}}}%
%BeginExpansion
\phantom{-}%
%EndExpansion
\psi_{Rk\downarrow}^{%
%TCIMACRO{\TeXButton{t}{\phantom{\dagger}}}%
%BeginExpansion
\phantom{\dagger}%
%EndExpansion
}\left(  \varpi\right)  \\%
%TCIMACRO{\TeXButton{-}{\phantom{-}}}%
%BeginExpansion
\phantom{-}%
%EndExpansion
\psi_{L\bar{k}\downarrow}^{\dagger}\left(  \bar{\varpi}\right)  \\
-\psi_{L\bar{k}\uparrow}^{\dagger}\left(  \bar{\varpi}\right)
\end{pmatrix}
\label{eqn:spinor}%
\end{equation}
(that we present directly in Fourier space). Here $k$ is the reduced momentum
(after linearization was carried out) and
\begin{equation}
\varpi=\omega-\mu\label{eqn:shiftfreq}%
\end{equation}
is the shifted frequency corresponding to a time evolution given by $K$
(cf.~with the discussion of tunneling given in Ref.~%
%TCIMACRO{\TeXButton{mahan2000}{[\onlinecite{mahan2000}]}}%
%BeginExpansion
[\onlinecite{mahan2000}]%
%EndExpansion
); here the bars have the meaning of minus signs. Since we make explicit
distinction between chiralities, all the components of the spinor are
independent. Using this basis the Hamiltonian can be written in matrix form,%
\begin{equation}
K_{\mathrm{sc}}=\Psi_{kn\sigma}^{\dagger}\left(  \varpi\right)
\begin{bmatrix}
\xi_{k\sigma}\hat{\sigma}_{\sigma\tau}^{0} & -\hat{\Delta}_{\sigma\tau}^{%
%TCIMACRO{\TeXButton{t}{\phantom{\dagger}}}%
%BeginExpansion
\phantom{\dagger}%
%EndExpansion
}\\
-\hat{\Delta}_{\sigma\tau}^{\dagger} & -\xi_{k\bar{\sigma}}\hat{\sigma
}_{\sigma\tau}^{0}%
\end{bmatrix}
_{nm}\Psi_{km\tau}\left(  \varpi\right)
\end{equation}
Here we arranged the different components of the order parameter using the
following matrix notation:
\[
\hat{\Delta}=%
\begin{pmatrix}
\Delta_{\downarrow\uparrow} & \Delta_{\uparrow\uparrow}\\
\Delta_{\downarrow\downarrow} & \Delta_{\uparrow\downarrow}%
\end{pmatrix}
\equiv\Delta_{a}\hat{\sigma}^{a}=%
\begin{pmatrix}
\Delta_{0}+\Delta_{3} & \Delta_{1}-i\Delta_{2}\\
\Delta_{1}+i\Delta_{2} & \Delta_{0}-\Delta_{3}%
\end{pmatrix}
\]
We note that another convention, the one introduced in the work of Balian and
Werthamer,\cite{balian1963} is related to ours via $\hat{\Delta}_{\mathrm{BW}%
}=\hat{\Delta}\cdot(i\hat{\sigma}^{y})$; the difference is rooted in a
different definition of the spinor basis.

In the case of zero magnetic field, $K^{2}$ is block diagonal and one arrives
to a closed solution for the quasiparticle excitation
spectrum.\cite{anderson1975} In the presence of magnetic field the
calculations for the case of a general order parameter are more involved. We
adopt the convention of taking the quantization axis (\thinspace$\hat{z}%
$\thinspace) along the magnetic field direction and consider the cases of
triplet order parameters parallel or perpendicular to the field. In both of
these cases the Hamiltonian can be diagonalized via a canonical rotation
(\textit{i.e.}~a Bogoliubov-Valatin transformation) that proceeds in a
completely identical way to that of the conventional \textit{s}-wave case.
Following the analogy further, the local Green functions for the leads can be
written down immediately. For the case of a parallel order parameter
(\textit{i.e.}~$\Delta_{1}=\Delta_{2}=0$) the non-zero matrix elements of the
advanced and retarded Green functions are:
\begin{align}
g_{11}^{r,a} &  =g_{33}^{r,a}=\frac{2}{w}\frac{-\left(  \varpi+h\pm
i\eta\right)  }{\sqrt{\left\vert \Delta_{\downarrow\uparrow}\right\vert
^{2}-\left(  \varpi+h\pm i\eta\right)  ^{2}}}\\
g_{13}^{r,a} &  =\left[  g_{31}^{r,a}\right]  =\frac{2}{w}\frac{\Delta
_{\downarrow\uparrow}^{\left[  \ast\right]  }}{\sqrt{\left\vert \Delta
_{\downarrow\uparrow}\right\vert ^{2}-\left(  \varpi+h\pm i\eta\right)  ^{2}}%
}\\
g_{22}^{r,a} &  =g_{44}^{r,a}=\frac{2}{w}\frac{-\left(  \varpi-h\pm
i\eta\right)  }{\sqrt{\left\vert \Delta_{\uparrow\downarrow}\right\vert
^{2}-\left(  \varpi-h\pm i\eta\right)  ^{2}}}\\
g_{24}^{r,a} &  =\left[  g_{42}^{r,a}\right]  =\frac{2}{w}\frac{\Delta
_{\uparrow\downarrow}^{\left[  \ast\right]  }}{\sqrt{\left\vert \Delta
_{\uparrow\downarrow}\right\vert ^{2}-\left(  \varpi-h\pm i\eta\right)  ^{2}}}%
\end{align}
with the upper (lower) sign corresponding to the retarded (advanced) ones.
Here $w=4v_{\mathrm{F}}$ is an energy scale related to the Fermi velocity (or
equivalently to the normal density of states at the Fermi level) and $\eta$ is
a positive infinitesimal that regularizes the Green functions (sometimes kept
finite to model the inelastic relaxation processes inside the leads).
Analogously for the case of a perpendicular order parameter (\textit{i.e.}%
~$\Delta_{0}=\Delta_{3}=0$), the non-zero matrix elements of the advanced and
retarded Green functions are this time:
\begin{align}
g_{11}^{r,a} &  =g_{44}^{r,a}=\frac{2}{w}\frac{-\left(  \varpi\pm
i\eta\right)  }{\sqrt{\left\vert \Delta_{\uparrow\uparrow}\right\vert
^{2}-\left(  \varpi\pm i\eta\right)  ^{2}}}\\
g_{14}^{r,a} &  =\left[  g_{41}^{r,a}\right]  =\frac{2}{w}\frac{\Delta
_{\uparrow\uparrow}^{\left[  \ast\right]  }}{\sqrt{\left\vert \Delta
_{\uparrow\uparrow}\right\vert ^{2}-\left(  \varpi\pm i\eta\right)  ^{2}}}\\
g_{22}^{r,a} &  =g_{33}^{r,a}=\frac{2}{w}\frac{-\left(  \varpi\pm
i\eta\right)  }{\sqrt{\left\vert \Delta_{\downarrow\downarrow}\right\vert
^{2}-\left(  \varpi\pm i\eta\right)  ^{2}}}\\
g_{23}^{r,a} &  =\left[  g_{32}^{r,a}\right]  =\frac{2}{w}\frac{\Delta
_{\downarrow\downarrow}^{\left[  \ast\right]  }}{\sqrt{\left\vert
\Delta_{\downarrow\downarrow}\right\vert ^{2}-\left(  \varpi\pm i\eta\right)
^{2}}}%
\end{align}
The non-equilibrium formalism that we seek to implement in order to access the
full I-V characteristics for arbitrary finite voltages, requires the
introduction of one more linearly independent Green function. From the
expressions for the retarded and advanced functions and using the assumption
of thermal equilibrium of the leads, we can construct immediately the
so-called Keldysh component\cite{keldysh1965} of the local lead Green
function: $g_{ij}^{k}=\left(  g_{ij}^{r}-g_{ij}^{a}\right)  \tanh\left(
\varpi/2T\right)  $.

Before proceeding, we stop to comment on the case of two normal leads. The
corresponding Green functions are obtained by using any of the two sets above
and taking the limit $\Delta_{a}\rightarrow0$ $\forall a$. In this case it is
a simple exercise to derive from Eq.~(\ref{eq:currentdef}) the well-known
expression for the conductance of an \textrm{N-N} junction:%
\begin{equation}
G_{\mathrm{NN}}=\frac{e^{2}}{\pi\hbar}\alpha\quad\text{with\quad}\alpha
=\frac{4t^{2}}{\left(  1+t^{2}\right)  ^{2}}~\text{,}%
\end{equation}
where we reintroduced Plank's constant and measured $t$ in units of $w$. This
expression was first derived by Landauer and later extended and generalized in
the works of B\"{u}ttiker, Imry and others.\cite{imry1999} The constant
$\alpha$ is called the \textit{channel transparency} and takes values in the
interval $\left[  0,1\right]  $. Now we return to the case when at least one
of the two leads is superconducting.

Given the local Green functions and the tunneling Hamiltonian, the simplest
way to proceed in order to compute the characteristics of a junction is to use
linear response and perturbation theory.\cite{wilkins1969,mahan2000} A more
rigorous approach should make use of non-equilibrium Green functions and treat
the tunneling term to all orders. This, to be able to calculate the full I-V
line and give a quantitative account of its sub-gap structure even in the
ballistic limit (\textit{i.e.}~for $\alpha\rightarrow1$). One past
implementation of this program made a clever use of the non-equilibrium Dyson
equations and reduced the problem to the solution of a set of linear recursion
relations.\cite{cuevas1996} Here we do not want to restrict ourselves to the
\textit{s}-wave case and to zero temperature and fields, we shall take then a
different route. We treat the local action and Green functions directly as
matrices in order to gain the convenience of a simpler implementation of
multiband multicomponent spinors and deal with them numerically.

\begin{figure}
[h]
\begin{center}
\includegraphics[
height=2.4396in,
width=2.623in
]%
{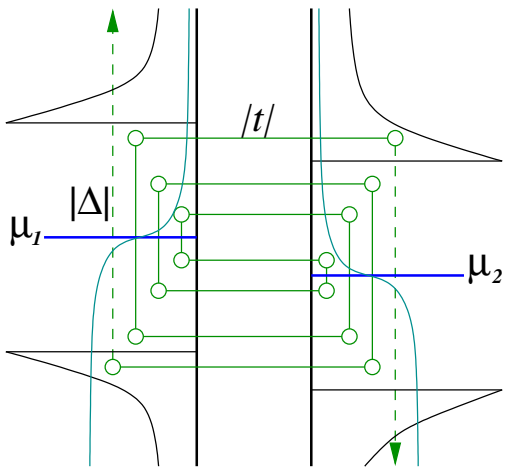}%
\caption{Depiction of a set of frequencies involved in a multiple coherent
tunneling process (the vertical axis corresponds to frequencies). The
horizontal lines correspond to frequency-conserving tunneling processes that
take place with amplitude $\left\vert t\right\vert $, while the vertical lines
correspond to electron (hole) pair creation or destruction processes that
occur with amplitude $\left\vert \Delta\right\vert $ (the superconducting gap
is taken in this figure to have the same magnitude in the two sides of the
junction). The chemical potentials, thermal distributions and superconducting
quasiparticle densities of states in the leads are schematically indicated.
The lowest order nonzero contribution to the tunneling is depicted. The dashed
lines and arrows indicate that the chain of interrelated higher order
processes continues \textit{ad infinitum}.}%
\label{andreev}%
\end{center}
\end{figure}

We notice that the lead Green functions can be inverted in close analytical
form to obtain the corresponding local lead actions. This procedure implies
the assumption of fast relaxation rates,\cite{wolf1989} which is consistent
with the point-contact geometry of the junction. Namely, working in a
Keldysh-extended Nambu-Eliashberg spinor basis of symmetric and antisymmetric
combinations of forward and backward time paths (see Ref.~%
%TCIMACRO{\TeXButton{keldysh1965}{[\onlinecite{keldysh1965}]}}%
%BeginExpansion
[\onlinecite{keldysh1965}]%
%EndExpansion
), the local action for a single lead can be written as
\begin{equation}
S_{\ell}=\int\frac{d\omega}{2\pi}\Psi_{\kappa n,\ell}^{\dagger}\left[
\mathcal{A}_{\ell}\right]  _{\kappa n,\kappa^{\prime}n^{\prime}}\Psi
_{\kappa^{\prime}n^{\prime},\ell}^{%
%TCIMACRO{\TeXButton{t}{\phantom{\dagger}}}%
%BeginExpansion
\phantom{\dagger}%
%EndExpansion
}%
\end{equation}
where $n$ labels the different components of the four-spinors as introduced in
Eq.~(\ref{eqn:spinor}) and $\kappa$ is the index for the two (symmetric and
antisymmetric) Keldysh components. The matrix representation of the spinorial
action density is given by
\begin{equation}
\mathcal{A}_{\ell}\equiv%
\begin{pmatrix}
\hat{0} & \hat{g}^{a}\\
\hat{g}^{r} & \hat{g}^{k}%
\end{pmatrix}
^{-1}=%
\begin{pmatrix}
-\left[  \hat{g}^{r}\right]  ^{-1}\hat{g}^{k}\left[  \hat{g}^{a}\right]  ^{-1}
& \left[  \hat{g}^{r}\right]  ^{-1}\\
\left[  \hat{g}^{a}\right]  ^{-1} & \hat{0}%
\end{pmatrix}
\end{equation}
where $\hat{g}^{r,a,k}$ are the matrices in the four-spinor basis whose
nonzero components were given above (for the two orientations of the order
parameter that we will consider, the inverses of $\hat{g}^{r,a}$ are easy to
calculate in closed form). By combining these actions and the spinorial matrix
representation of the tunneling Hamiltonian ($\mathcal{H}_{\mathrm{tun}}$)
written in a two-lead Keldysh-extended Nambu-Eliashberg spinor basis, one can
ensemble the full non-equilibrium action matrix density for the junction%
\begin{equation}
\mathcal{A}=\left[  \mathcal{A}_{\ell=1}\oplus\mathcal{A}_{\ell=2}\right]
-\mathcal{H}_{\mathrm{tun}}~\text{.}\label{eqn:fullaction}%
\end{equation}
While carrying out this construction, special attention must be paid to the
fact that the shifted frequencies [see Eq.~(\ref{eqn:shiftfreq})] will have
different reference levels when there is a relative bias applied to the leads.
Positively- and negatively-shifted-frequencies in each lead are related by the
coherent pairing processes in the superconductors; this is reflected on the
choice of frequency pairs in the spinor basis. When two superconductors with
different chemical potentials are put into contact, the tunneling Hamiltonian
connects real (\textit{i.e.}~\textit{unshifted}) frequencies. Thus, at finite
voltages, pairing and tunneling together create an infinite set of related
frequencies that is at the heart of the multiparticle tunneling processes
mediated by the so-called Andreev reflections; this is illustrated in
Fig.~\ref{andreev}.%

%TCIMACRO{\FRAME{fhFU}{2.623in}{2.4396in}{0pt}{\Qcb{Depiction of a set of
%frequencies involved in a multiple coherent tunneling process (the vertical
%axis corresponds to frequencies). The horizontal lines correspond to
%frequency-conserving tunneling processes that take place with amplitude
%$\left\vert t\right\vert $, while the vertical lines correspond to electron
%(hole) pair creation or destruction processes that occur with amplitude
%$\left\vert \Delta\right\vert $ (the superconducting gap is taken in this
%figure to have the same magnitude in the two sides of the junction). The
%chemical potentials, thermal distributions and superconducting quasiparticle
%densities of states in the leads are schematically indicated. The lowest order
%nonzero contribution to the tunneling is depicted. The dashed lines and arrows
%indicate that the chain of interrelated higher order processes continues
%\textit{ad infinitum}.}}{\Qlb{andreev}}{andreev.eps}%
%maintain-aspect-ratio TRUE;  display "ICON";  valid_file "F";  width 2.623in;
%height 2.4396in;  depth 0pt;  original-width 2.5815in;
%original-height 2.3981in;  cropleft "0";  croptop "1";  cropright "1";
%cropbottom "0";  filename 'andreev.eps';file-properties "XNPEU";}}}%
%BeginExpansion
%EndExpansion

To each value in the frequency window defined by the chemical potentials in
the two leads, one such set of `entangled' frequencies can be assigned. These
sets are independent and the action is block diagonal between different ones.
Discretizing the frequencies in this window automatically defines a
discretization of the `whole' frequency space. We proceed in this way and deal
with one such set of frequencies at a time. Since these sets are infinite, we
truncate their hierarchies at some distance from the central frequency window.
This is equivalent to introducing a \textit{soft} limit in the number of
allowed Andreev reflections: up to some fixed number ($N_{\mathrm{A}}$) they
are fully taken into account and then they are gradually suppressed until
twice that number is reached. This is a natural and consistent cut-off scheme
at any finite voltage, since the presence of a growing frequency denominator
makes the higher contributions less and less important regardless of the value
of $\alpha$. It is also clear what the limitations of the approach are: as the
difference in chemical potentials decreases, the denominators grow more and
more slowly and a larger number of Andreev reflections is required in order to
achieve the same accuracy.

The implementation of the described scheme imports one more complication. In
the spinor basis we adopted, only chirality conserving tunneling processes can
be written in matrix form. To overcome this problem we introduce a second
\textit{mirrored} spinor basis with the chiralities inverted (as in
Eq.~(\ref{eqn:spinor}) but interchanging $R\leftrightarrow L$). Using two
copies of the spinor space the full tunneling Hamiltonian (that is, including
chirality non-conserving processes) can be written as a matrix and the
\textit{whole} frequency space for \textit{both} chiral species is considered
(let us stress that no Hilbert space doubling takes place). Inverting the
action matrix density thus constructed [Eq.~(\ref{eqn:fullaction})], using
standard numerical methods, we can obtain frequency densities for the
different current harmonics (constructed out of the Keldysh components of
lead-mixing Green functions). Here we will concentrate on the dc component.
Finally, the current is computed integrating its density over the full
frequency axis,%
\begin{equation}
I=\frac{et}{2i}\sum_{\sigma}\int\frac{d\omega}{2\pi}\left\langle
\psi_{2,\sigma}^{\dagger}\psi_{1,\sigma}^{%
%TCIMACRO{\TeXButton{t}{\phantom{\dagger}}}%
%BeginExpansion
\phantom{\dagger}%
%EndExpansion
}-\psi_{1,\sigma}^{\dagger}\psi_{2,\sigma}^{%
%TCIMACRO{\TeXButton{t}{\phantom{\dagger}}}%
%BeginExpansion
\phantom{\dagger}%
%EndExpansion
}\right\rangle _{\mathrm{kel}}~\text{.}%
\end{equation}

The practical implementation of this numerical scheme is straightforward and
allows one to consider the (combined) effects of finite temperature, applied
magnetic fields, contact potentials in the junction, spin-flip tunneling or
spin-flip scattering processes\cite{grimaldi1997} in the leads. It is also
possible to compute the ac response. Here our primary interest is in comparing
singlet and triplet superconductor junctions and how they respond differently
in the presence of an external field or with temperature; other additional
complications will be discussed elsewhere. Even though our numerical scheme is
not well suited for studying the limit $V\rightarrow0$, in particular the
combination $\alpha\thicksim1$ with $V\thicksim0$ is the computationally most
expensive one, that limit can nevertheless be approached analytically. On the
other hand, all other regimes can be solved with modest computational effort
and the algorithm is quite easily parallelizable. Even more, we shall show
that the finite bias features are the ones that might provide useful
signatures to further establish the spin-triple scenario.

\section{Tunneling Characteristics\label{sec:tunchar}}

To discuss and compare the I-V characteristics for different types of
junctions, we choose some convenient set of parameters that clearly display
the different features. For the tunneling overlap integral we choose the
values $t=0.2$ and $t=0.5$ (that correspond, in the notation of Ref.~%
%TCIMACRO{\TeXButton{blonder1982}{[\onlinecite{blonder1982}]}}%
%BeginExpansion
[\onlinecite{blonder1982}]%
%EndExpansion
, to $\alpha\simeq0.15$ or $Z=2.4$ and to $\alpha=0.64$ or $Z=0.75$,
respectively), and when there is a magnetic field we fix its value to $h=0.2$
in units of $\Delta$ (by $\Delta$ we mean the magnitude of the singlet gap,
$\Delta_{0}$, or of the triplet vector order parameter depending on the case;
notice we absorbed Bohr's magneton and the gyromagnetic factor in the
definition of the magnetic field). These values are larger than, for instance,
those in the most typical STM tunneling experiments, except for the ones
engineered expressly to seek for large values of $\alpha$,\cite{scheer1998}
but have the virtue of making evident the different features, including the
Andreev gap structure (see below). We show, except when indicated, curves for
the dc response in the limit of vanishing temperatures. For the truncation
procedure we have taken $N_{\mathrm{A}}=3$ and $N_{\mathrm{A}}=5$ for the
cases of $t=0.2$ and $t=0.5$, respectively (and verified that larger values
produce, given the set of parameters chosen, identical curves). The
discretization used on the horizontal axis is better than $\delta
V=0.025$~$\Delta/e$ in all the cases.%

%TCIMACRO{\FRAME{ftbFU}{3.3883in}{3.282in}{0pt}{\Qcb{Zero temperature I-V
%characteristics of normal-superconductor junctions for both spin-singlet and
%spin-triplet paring. (a) N-S junctions for $t=0.2$ (lower curves) with and
%without applied magnetic field (dashed and solid line respectively, $h=0.2$)
%and $t=0.5$ (upper curve, solid line only since the effect of field is smaller
%and not displayed); the curves are vertically displaced for clarity. (b) N-T
%junctions for $t=0.2$ (upper curve) with and without applied magnetic field
%(dashed and solid line respectively, $h=0.2$) and $t=0.5$ (lower curve, solid
%line only).}}{\Qlb{nsgraph2}}{nsgraph2.eps}%
%width 3.3883in;  height 3.282in;  depth 0pt;  original-width 5.0004in;
%original-height 3.4938in;  cropleft "0";  croptop "1";  cropright "1";
%cropbottom "0";  filename 'NSgraph2.eps';file-properties "XNPEU";}}}%
%BeginExpansion
\begin{figure}
[tb]
\begin{center}
\includegraphics[
height=3.282in,
width=3.3883in
]%
{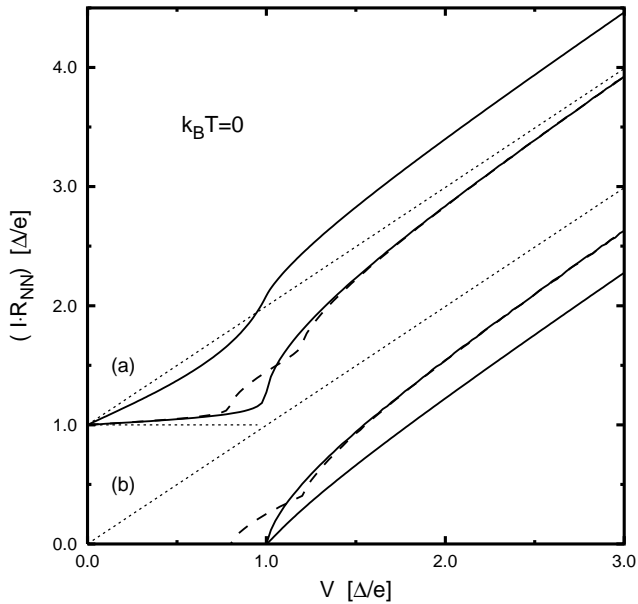}%
\caption{Zero temperature I-V characteristics of normal-superconductor
junctions for both spin-singlet and spin-triplet paring. (a) N-S junctions for
$t=0.2$ (lower curves) with and without applied magnetic field (dashed and
solid line respectively, $h=0.2$) and $t=0.5$ (upper curve, solid line only
since the effect of field is smaller and not displayed); the curves are
vertically displaced for clarity. (b) N-T junctions for $t=0.2$ (upper curve)
with and without applied magnetic field (dashed and solid line respectively,
$h=0.2$) and $t=0.5$ (lower curve, solid line only).}%
\label{nsgraph2}%
\end{center}
\end{figure}
%EndExpansion

We review now the different pairing-symmetry scenarios. Let us start with the
case of normal-metal--superconductor junctions. We show in Fig.~\ref{nsgraph2}%
-(a) typical curves for an N-S junction (\textit{i.e.}~a point-contact
junction between a normal metal and a conventional singlet-paring
superconductor). The diagonal straight line is the N-N characteristics given
as a reference. The solid lines correspond to the N-S junction in zero field
and the dashed line is for one of the junctions (the less transparent one) in
the presence of a magnetic field. The effect of the magnetic field is to
produce what would be seen as a Zeeman splitting of the differential
conductance peak (\textit{i.e.}~the peak in the curve of $dI/dV$
\textit{vs}.~$V$). Notice the sub-gap shoulder on the I-V curve when
$eV<\Delta$ (for instance in the zero field case); its origin is in the
coherent Andreev processes that take place at the junction contact. Next we
show in Fig.~\ref{nsgraph2}-(b) a typical curve this time for what we call an
N-T junction (\textit{i.e.}~a junction between a normal metal and an
unconventional triplet-pairing superconductor). The solid lines correspond to
the N-T junction in zero field and the dashed line is for the $t=0.2$ junction
when in the presence of a magnetic field that is aligned with the vector order
parameter $\vec{\Delta}$. If one considers a magnetic field that is
perpendicular to the order parameter ($\vec{h}\perp\vec{\Delta}$), one finds
it has no effect on the I-V characteristic that remains identical to the one
for the zero field case (remark that in the case of the N-S junction the
orientation of the field was immaterial). Notice also the absence of a sub-gap
shoulder on the I-V curve. This absence is caused by the odd real-space
symmetry of the superconductor (\textit{p}-wave paring): Andreev processes
with opposite chiralities interfere destructively and exactly cancel each
other. As a result, the curves are exactly identical to those computed with a
semiconducting band model that ignores Andreev scattering (to be contrasted
with the non-trivial results in this respect that will be shown momentarily
for junctions involving two different-symmetry superconductors).%

%TCIMACRO{\FRAME{ftbFU}{3.5276in}{2.4734in}{0pt}{\Qcb{Zero temperature I-V
%characteristics of S-T junctions with and without magnetic field (dashed and
%solid lines respectively, $h=0.2$). The lower solid and dashed curves are for
%$t=0.2$ (without and with magnetic field) and the upper solid curve is for
%$t=0.5$. The dotted lines are (i) the straight unitary slope line for the
%reference N-N characteristics and (ii) the S-S characteristics for
%similar-parameters junctions (upper $t=0.5$ and lower $t=0.2$). The inset
%shows finite temperature characteristics of S-S junctions with different
%values of the left and right gap amplitudes. The dotted line is the reference
%N-N characteristics and the dashed line is the curve for $\Delta_{2}%
%=\Delta_{1}$. The solid lines correspond to $\Delta_{2}\neq\Delta_{1}$
%($\Delta_{2}=3~\Delta_{1}$ for the line closest to the dashed one and
%$\Delta_{2}=37/3~\Delta_{1}$ for the other one).}}{\Qlb{stgraph2}%
%}{stgraph2.eps}{\special{ language "Scientific Word";  type "GRAPHIC";
%width 3.5276in;  height 2.4734in;  depth 0pt;  original-width 5.0004in;
%original-height 3.4938in;  cropleft "0";  croptop "1";  cropright "1";
%cropbottom "0";  filename 'STgraph2.eps';file-properties "XNPEU";}}}%
%BeginExpansion
\begin{figure}
[tb]
\begin{center}
\includegraphics[
height=2.4734in,
width=3.5276in
]%
{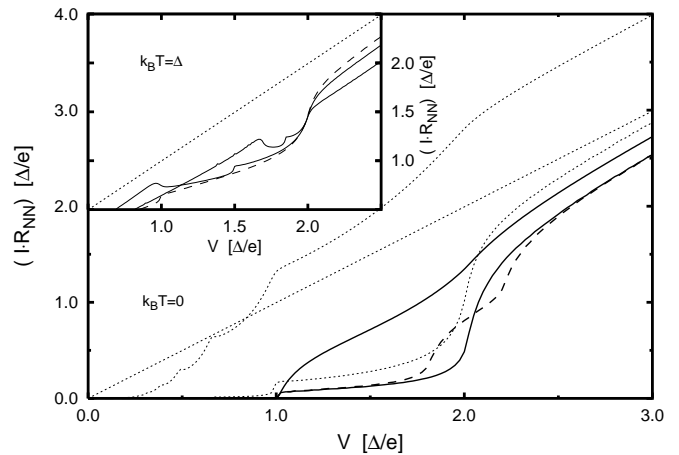}%
\caption{Zero temperature I-V characteristics of S-T junctions with and
without magnetic field (dashed and solid lines respectively, $h=0.2$). The
lower solid and dashed curves are for $t=0.2$ (without and with magnetic
field) and the upper solid curve is for $t=0.5$. The dotted lines are (i) the
straight unitary slope line for the reference N-N characteristics and (ii) the
S-S characteristics for similar-parameters junctions (upper $t=0.5$ and lower
$t=0.2$). The inset shows finite temperature characteristics of S-S junctions
with different values of the left and right gap amplitudes. The dotted line is
the reference N-N characteristics and the dashed line is the curve for
$\Delta_{2}=\Delta_{1}$. The solid lines correspond to $\Delta_{2}\neq
\Delta_{1}$ ($\Delta_{2}=3~\Delta_{1}$ for the line closest to the dashed one
and $\Delta_{2}=37/3~\Delta_{1}$ for the other one).}%
\label{stgraph2}%
\end{center}
\end{figure}
%EndExpansion

Let us now turn to examine the case of junctions in which both their sides are
superconducting. In Fig.~\ref{stgraph2} we display typical curves for S-S
junctions (both the sides are conventional spin-singlet superconductors) and
S-T junctions (one of the sides is a spin-triplet superconductor). The
straight dotted line is the N-N characteristics --~taken as a reference, same
as before. The remaining dotted lines are the I-V curves of S-S junctions that
show all the standard features already well documented in the
literature.\cite{octavio1983,cuevas1996} For the purpose of later comparison,
we remark here the sizeable currents for voltages $eV>2\Delta$ (the value of
the gap is taken to be the same on both sides of the junction), and the
`sub-gap' shoulder with Andreev steps at $eV=2\Delta/n$ (with $n=1,2,3,\ldots
$). We also remind the reader that this curve is, when orbital effects can be
ignored, not sensitive to applied magnetic fields. The remaining curves
(dashed and solid lines) correspond to S-T junctions with different tunneling
matrix element strengths and with and without magnetic field (respectively).
The solid lines are insensitive to the orientation of the vector order
parameter on the triple-pairing side of the junctions, and the current
amplitude is found to be systematically smaller than in the case of the
respective S-S junctions. Remarkably, the `sub-gap' structure shows only two
steps (at voltages given by $n=1,2$) and the current is zero when $eV<\Delta$
(if the magnitudes of the gaps in the spin-singlet and spin-triplet sides of
the junction are different, then the zero current condition is $eV<\Delta
_{\mathrm{Triplet}}$, where $\Delta_{\mathrm{Triplet}}$ is the magnitude of
the vector order parameter on the spin-triplet side of the junction).
Concerning the effects of an applied magnetic field, the curves remain
unchanged if the field is applied parallel to the direction of the
vector-order-parameter, but show instead a Zeeman effect if the field is
perpendicular to it (dashed line). This is in contrast with the case of N-T
junctions, for which the Zeeman effect is expected for fields $\vec
{h}\parallel\vec{\Delta}$.

In the figure inset we display curves for S-S junctions at finite
temperatures. In order to render the different features simultaneously
visible, we `push' the temperature to be equal to $\Delta$ (with
$k_{\mathrm{B}}=1$). Let us denote as $\Delta$ the average gap value
($2\Delta=\Delta_{1}+\Delta_{2}$) that we keep using to define the unit in
which we measure the voltage, normalized current, etc. The dashed line is the
finite temperature I-V for the $t=0.2$ junction between two identical
superconductors ($\Delta_{2}=\Delta_{1}$), whereas the solid lines correspond
to similar junctions with $\Delta_{2}=3~\Delta_{1}$ or $\Delta_{2}%
=37/3~\Delta_{1}$ and the same transparency (during this discussion we assume
$\Delta_{2}\geq\Delta_{1}$). Besides the standard quasiparticle tunneling
threshold ($eV=2\Delta$) one clearly sees in both solid curves the step at
$eV=\Delta_{2}$ corresponding to the first term of the so-called \textit{even
series} ($eV=2\Delta_{\ell}/m$ with $\ell=1,2$ and $m=2,4,\ldots$). The steps
corresponding to the \textit{odd series} ($eV=2\Delta/m$ with $m=3,5,\ldots$)
are not visible in this plot, but we verified that we observe them at low
temperatures in junctions with $\Delta_{2}\gtrsim\Delta_{1}$. This current
step structure lacks temperature or magnetic field dependence (in accordance
with experimental observations). The other prominent feature of the solid
curves is the rounded cusp at $eV\approx\Delta_{2}-\Delta_{1}$. This is a
thermally activated feature that appears when the upper gap edges at both
sides of the junctions are aligned. The rounding of the cusp has similar
origin as the rounding of the quasiparticle threshold, both are due to higher
order multiparticle processes. One new observation that we made is that the
position of the thermal cusp is not exactly $\Delta_{2}-\Delta_{1}$, but it is
shifted towards lower voltages. This is again a result of taking into account
higher order Andreev processes and is made more evident by our choice of
parameters; for lower temperatures and less transparent junctions this
correction is typically very small. We also find that the steps of both the
even and the odd series that fall to the left of the thermal cusp are usually
washed away by its tail; this we find to be consistent with available
experimental results. We expect the different detailed features corresponding
to dissimilar gaps and finite temperatures to be accessible to current
state-of-the-art experiments, we will comment on that in the next
section.\bigskip

As an aside, let us comment on the effect of local contact potential terms in
the tunneling matrix. They suppress uniformly the dc current amplitude but
have no other effect on the shape of the I-V characteristics. This is true
regardless of the pairing symmetry of the superconductors forming the
junction, in particular they do not cause the appearance of mid-gap states in
the case of triplet pairing (for neither the N-T or S-T junctions nor for the
T-T junctions discussed below).\bigskip%

%TCIMACRO{\FRAME{ftbFU}{3.5276in}{2.4734in}{0pt}{\Qcb{Zero temperature I-V
%characteristics of T-T junctions. The lower curve corresponds to $t=0.5$ and
%the upper one to $t=0.2$. The dotted lines is the reference N-N
%characteristics. The inset shows finite temperature characteristics of T-T
%junctions with different values of the left and right gap amplitudes. We used
%the same values choices as in the inset of Fig.~\ref{stgraph2}. Namely, the
%dotted line is the N-N characteristics and the dashed line corresponds to
%$\Delta_{2}=\Delta_{1}$ while the solid lines are for $\Delta_{2}\neq
%\Delta_{1}$($\Delta_{2}=3~\Delta_{1}$ for the line closest to the dashed one
%and $\Delta_{2}=37/3~\Delta_{1}$ for the other one).}}{\Qlb{ttgraph2}%
%}{ttgraph2.eps}{\special{ language "Scientific Word";  type "GRAPHIC";
%width 3.5276in;  height 2.4734in;  depth 0pt;  original-width 5.0004in;
%original-height 3.4938in;  cropleft "0";  croptop "1";  cropright "1";
%cropbottom "0";  filename 'TTgraph2.eps';file-properties "XNPEU";}}}%
%BeginExpansion
\begin{figure}
[tb]
\begin{center}
\includegraphics[
height=2.4734in,
width=3.5276in
]%
{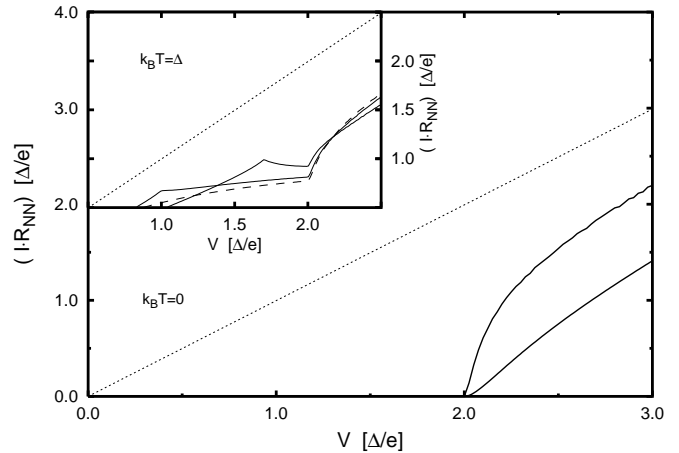}%
\caption{Zero temperature I-V characteristics of T-T junctions. The lower
curve corresponds to $t=0.5$ and the upper one to $t=0.2$. The dotted lines is
the reference N-N characteristics. The inset shows finite temperature
characteristics of T-T junctions with different values of the left and right
gap amplitudes. We used the same values choices as in the inset of
Fig.~\ref{stgraph2}. Namely, the dotted line is the N-N characteristics and
the dashed line corresponds to $\Delta_{2}=\Delta_{1}$ while the solid lines
are for $\Delta_{2}\neq\Delta_{1}$($\Delta_{2}=3~\Delta_{1}$ for the line
closest to the dashed one and $\Delta_{2}=37/3~\Delta_{1}$ for the other
one).}%
\label{ttgraph2}%
\end{center}
\end{figure}
%EndExpansion

Finally, the only remaining case to consider is that of junctions in which
both sides are spin-triplet superconductors. Such a case is exemplified in the
curves of Fig.~\ref{ttgraph2}. Andreev processes with triplet symmetric
paring, for tunneling through a single-mode contact, interfere destructively
and the current remains zero up to voltages larger than $eV=2\Delta$, when
quasiparticle tunneling becomes allowed. The lower solid line corresponds to
$t=0.5$ and the upper one to $t=0.2$, the inversion of the order is due to the
fact that, for this range of parameters, the current at fixed voltage grows
more slowly than in the case of normal junctions (N-N). Similarly as in the
S-S case, both sides of the junction react identically to applied magnetic
fields and no net effects are therefore visible in the current-voltage characteristics.

The curves in the figure inset are the same as in the inset of the previous
figure but for the case of spin-triple pairing symmetry (on both sides of the
junction). Both the quasiparticle tunneling threshold and the thermal cusp
remain sharp since higher order processes interfere destructively and no
rounding takes place. Another consequence of this is that the position of the
thermal cusp is exactly $eV=\Delta_{2}-\Delta_{1}$ and no shift is observed.
The sub-gap part of the curves is smooth and its height and shape are governed
by the thermal excitations, the step structure of the even and odd series is
absent. Also in this case magnetic fields have no direct effects.

\section{Experimental consequences\label{sec:experiments}}

A first set of applications concerns atomic contacts. For the case of normal
or singlet superconducting leads with identical gaps, and zero temperature and
magnetic field, our results are in full agreement with the previous studies of
such systems.\cite{cuevas1996} For the case of two different gaps shown in the
inset of Fig.~\ref{stgraph2}, our theory correctly reproduces the different
steps as discussed in the previous section. Another prominent feature of such
curves is the thermally activated rounded cusp at $eV\approx\Delta_{2}%
-\Delta_{1}$. The rounding is due to higher order multiparticle processes; and
a new prediction is that the position of the thermal cusp is not exactly
$\Delta_{2}-\Delta_{1}$, but it is shifted towards lower voltages. Such a
shift from the naive \textquotedblleft density of states\textquotedblright%
\ answer could in principle be checked directly in atomic contacts. We also
find that the steps of both the even and the odd series that fall to the left
of the thermal cusp are usually washed away by its tail; this we find to be
consistent with available experimental results.\cite{wolf1989} Such features
corresponding to dissimilar gaps and finite temperatures should be accessible
to current state-of-the-art experiments. For instance, the experiments of
Ref.~%
%TCIMACRO{\TeXButton{scheer1998}{[\onlinecite{scheer1998}]} }%
%BeginExpansion
[\onlinecite{scheer1998}]
%EndExpansion
could be attempted using a \textrm{Pb} STM tip as before but to prove into a
\textrm{Mn}-doped \textrm{Pb} sample. \textrm{Mn} will act as a magnetic
impurity and decrease the value of the gap, such a setup would correspond to
the situation of similar but not identical gap parameters with a doping
controllable difference; it would be a way to try to observe the `splitting'
of the even series in single-point contacts. Other setups could be envisaged
based also on STM techniques or on pressed crossed wires.\bigskip

The main application of our results, however, concerns the use of tunneling
with triplet superconductors.\cite{bolech2004} In that case the most direct
experimental realization is organic superconductors.\cite{chemicalreview} The
experiments show that for magnetic fields along the direction of the
conducting chains ($\mathbf{a}$ crystalline-axis) the upper critical field is
paramagnetically limited. If such systems are indeed triplet superconductors,
this would correspond, following our notations, to a vector order parameter
aligned with the field ($\vec{h}\parallel\vec{\Delta}$).\cite{lebed2000} With
this geometry a Zeeman splitting of the differential conductance peak, similar
to that in conventional superconductors, should be observed in a tunneling
experiment. As the field is rotated the splitting would be suppressed and for
a magnetic field oriented parallel to the $\mathbf{b}^{\prime}$
crystalline-axis there should be no Zeeman effect (accompanied by the
possibility of applying large fields that are not paramagnetically limited).
The disappearance of splitting even as the field is being increased would
constitute a clear signature of spin-triplet superconductivity. The main
difficulties of such an experiment would be the set up of point-contacts and
the resolution required to observe the Zeeman effect. On the first point one
possibility would be to use STM setups with `thin tips'. On the second point,
since the critical temperature of these organic salts is relatively low, the
experiment could be done with moderate fields that would produce splittings
that are a substantial fraction of the superconducting gap. The linearity of
the magnetic field dependence of these splittings, a signature of the Zeeman
effect, could be accurately established using Fourier analysis
techniques.\cite{hertel1985}

Similarly as in the case of N-T junctions, we can envisage using the Zeeman
response of S-T junctions as a direct probe for spin-triplet order. If, for
instance, a magnetic field is applied along the $\mathbf{b}^{\prime}$
crystalline-axis of \textrm{(TMTSF)}$_{2}$\textrm{PF}$_{6}$, we predict a
Zeeman splitting of the main differential conductance peak. This would also
constitute a clear sign of unconventional superconductivity since such an
effect does not take place for standard BCS superconductors. The
$\mathbf{b}^{\prime}$ direction is the one on which the upper critical field
is not paramagnetically limited, so relatively large fields could be applied
in order to obtain a clear signal (as the field alignment changes the
splitting disappears). To afford large fields one would need to use in the
`conventional' side of the junction a compound with a relatively high critical
temperature (as compared with that of \textrm{(TMTSF)}$_{2}$\textrm{PF}$_{6}%
$). In this respect one has the bonus that, since the required setup should be
a point-contact, superconductivity might survive at the contact-neck region up
to fields much in excess of the bulk value of $H_{\mathrm{c2}}$ (rather
approaching $H_{\mathrm{p}}^{\mathrm{BCS}}$ for that
material).\cite{clogston1962,suderow2002} Another advantage in the
two-superconductor setup is that the levels of noise are usually
smaller,\cite{eskildsen2003} allowing a better definition of the differential
conductance signal from where the Zeeman splitting is going to be read off.

Similar considerations could also be made for those layered compounds that are
believed to be triplet superconductors.\cite{mineev2002,ishiguro2002} In that
case the critical magnetic field is not paramagnetically limited when the
applied field is oriented parallel to the superconducting planes. Among these
compounds \textrm{Sr}$_{2}$\textrm{RuO}$_{4}$ is the best studied one so far,
but only few tunneling experiments were
performed,\cite{jin2000,laube2000,mao2001,upward2002,kugler2003} and none so
far with high resolution and in the presence of an applied external magnetic
field (see though Refs.~%
%TCIMACRO{\TeXButton{mao2001,sumiyama2002,upward2002}{[\onlinecite
%{mao2001,sumiyama2002,upward2002}]}}%
%BeginExpansion
[\onlinecite{mao2001,sumiyama2002,upward2002}]%
%EndExpansion
). One of the conspicuous features observed in some of these experiments is
the presence of a `zero bias anomaly' (ZBA) in the differential conductance.
Its explanation is still a matter of debate, but seems to require extended
contact interfaces and momentum dependent order parameters (to include the
effect of `zero energy states' at the interfaces, extensions to our scheme
would be required, possibly incorporating certain aspects of those
calculations already done for planar
junctions\cite{yamashiro1997,honerkamp1998,asano2003}). Our general findings
about the effect of magnetic fields should, however, apply, since they refer
to effects to be measured at voltages of the order of the superconducting gap.
It is intriguing to notice that in the `point-contact' experiment of Ref.~%
%TCIMACRO{\TeXButton{laube2000}{[\onlinecite{laube2000}]} }%
%BeginExpansion
[\onlinecite{laube2000}]
%EndExpansion
two types of spectra are measured: with and without ZBA in the differential
resistance. One might speculate that what changes between the different
samples should be no other thing than the effective size, potential barrier
and geometry of the contact (experiments in other compounds indicate that that
might be enough to give rise to zero voltage features; cf.~Ref.~%
%TCIMACRO{\TeXButton{sheet2004}{[\onlinecite{sheet2004}]}}%
%BeginExpansion
[\onlinecite{sheet2004}]%
%EndExpansion
). In particular, if that is the case, at least those $dV/dI$ curves with no
ZBA should, according to our calculations, show no Zeeman splitting when a
field is applied parallel to the \textrm{Ru-O }planes; in contrast to what is
expected for BCS superconductors. Two different groups reported that further
point contact and STM tunneling experiments on \textrm{Sr}$_{2}$%
\textrm{RuO}$_{4}$ in a magnetic field are
underway.\cite{laube2000,kambara2003}

\section{Summary and Closing Remarks\label{sec:summary}}

Summarizing, we have shown how the full I-V characteristics for point-contact
junctions can be accurately studied using a local action approach in the
context of the Keldysh formalism. Our formalism allows one to treat both
normal and superconducting (singlet and triplet) leads, and to take into
account effects of finite magnetic field and temperature. In particular we
have shown that the point-contact tunneling involving unconventional
superconductors with spin-triplet pairing displays interesting characteristic
features. Unlike the case of conventional superconductors, these show quite
different characteristics whether the junctions are planar\cite{sengupta2001}
or point-contact like.\cite{bolech2004} The Zeeman response to an external
magnetic field is such that it allows for the identification of triplet phases
and might be relevant for future experiments. The prediction of a truncated
sub-gap structure in point-contact S-T junctions is also very interesting, but
experiments to test this are much harder to carry out. That kind of detailed
experiments are, however, possible for conventional superconductors and we
believe they could be exploited to look for some as yet poorly tested
predictions of the theory. For instance, the experiments of Ref.~%
%TCIMACRO{\TeXButton{scheer1998}{[\onlinecite{scheer1998}]} }%
%BeginExpansion
[\onlinecite{scheer1998}]
%EndExpansion
could be attempted using a \textrm{Pb} STM tip as before but doping the sample
with \textrm{Mn}; it would be a way to try to observe the `splitting' of the
even series in single-point contacts.

Besides the different additional effects on the tunneling characteristic that
we discussed here (the effects of fields, temperature and contact potentials),
there are others that can also be easily taken into account like, for
instance, spin-flip tunneling processes or temperature gradients. These
effects will be relevant in the study of junctions involving ferromagnets (of
possible relevance in the context of spintronics; cf.~Ref.~%
%TCIMACRO{\TeXButton{lopez2003}{[\onlinecite{lopez2003}]}}%
%BeginExpansion
[\onlinecite{lopez2003}]%
%EndExpansion
) or in precision studies related to the renewed interest in the use of
microjunctions for \textit{on-chip} thermometry and eventually
cryogenics.\cite{pekola2004} A remaining challenge, however, is how to extend
our formalism seeking to include the physics responsible for producing ZBAs in
order to further our understanding of tunneling experiments in layered
unconventional superconductors and planar junctions. Such extensions shall
seek to describe not only the point-contact limit, but also the planar
interface case. This is one of the ingredients necessary for a realistic
description of layered materials like the ruthenates. In that respect the
interplay with the advances already made using semiclassical methods will
constitute not only a check of the approximations made in the latter but also
a way of finding efficient computational schemes for more complicated
scenarios that might incorporate, for instance, the two-band nature of certain
compounds. On the other hand, in the context of the quasi-one-dimensional
organic conductors, such developments should help to go beyond the
one-dimensional approximation.

\begin{acknowledgments}
We would like to thank \O .~Fischer, M.~Eskildsen, M.~Kugler and G.~Levy for
discussions about tunneling and STM. We also thank Y.~Maeno and M.~Sigrist for
discussions about tunneling into triplet superconductors and ruthenates in
particular. This work was done under the auspices of the Swiss National
Science Foundation and its MaNEP program.
\end{acknowledgments}

%\bibliographystyle{apsrev}
%\bibliography{strings,tunneling}

\end{document}